\documentclass[runningheads]{llncs}
\usepackage{graphicx}
\usepackage{xcolor}
\usepackage{cite}
\usepackage{amsmath,amssymb,amsfonts}
\usepackage{textcomp}
\usepackage{multirow}
\usepackage{marvosym}
\usepackage{graphicx}
\usepackage{newclude}
\usepackage{float}
\usepackage{multirow}
\usepackage{subfigure}

\usepackage[noend]{algpseudocode}
\usepackage{algorithmicx, algorithm}
\usepackage{amsfonts}
\usepackage{amsmath}

\begin{document}

\title{Topic-Grained Text Representation-based Model for Document Retrieval}

\author{\textbf{*}Mengxue Du, \textbf{*}Shasha Li, Jie Yu\textsuperscript{\Letter}, Jun Ma\textsuperscript{\Letter}, Bin Ji, Huijun Liu, Wuhang Lin \and Zibo Yi}

\authorrunning{M. Du et al.}

\institute{College of Computer, National University of Defense Technology,
Changsha, Hunan Province, China\\
\email{\{dumengxuenudt,shashali,yj,majun, jibin,liuhuijun,wuhanglin,yizibo14\}@nudt.edu.cn}}
\maketitle

\begin{abstract}
Document retrieval enables users to find their required documents accurately and quickly. To satisfy the requirement of retrieval efficiency, prevalent deep neural methods adopt a representation-based matching paradigm, which saves online matching time by pre-storing document representations offline. However, the above paradigm consumes vast local storage space, especially when storing the document as word-grained representations. To tackle this, we present \textbf{TGTR}, a \textbf{T}opic-\textbf{G}rained \textbf{T}ext \textbf{R}epresentation-based \textbf{Model} for document retrieval. Following the representation-based matching paradigm, TGTR stores the document representations offline to ensure retrieval efficiency, whereas it significantly reduces the storage requirements by using novel topic-grained representations rather than traditional word-grained. Experimental results demonstrate that compared to word-grained baselines, TGTR is consistently competitive with them on TREC CAR and MS MARCO in terms of retrieval accuracy, but it requires less than 1/10 of the storage space required by them. Moreover, TGTR overwhelmingly surpasses global-grained baselines in terms of retrieval accuracy.
\end{abstract}

\section{Introduction}\label{introduction}
Recently, deep learning based semantic representations have attracted much research attention and been widely used in the document retrieval field. Recent methods propose to fine-tune deep pre-trained language models (PLMs) such as BERT \cite{bert} to assess matching degrees of query-document pairs \cite{colbert,bert-qe,bert}. They achieve the state-of-the-art performance of the document retrieval task by concatenating query-document pair and feeding it into a PLM to calculate the matching degree. Unfortunately, despite these methods achieve great success, they come at a steep increase in time cost, which is unacceptable in practical application scenarios.

\begin{figure}
\centering
\includegraphics[width=\textwidth]{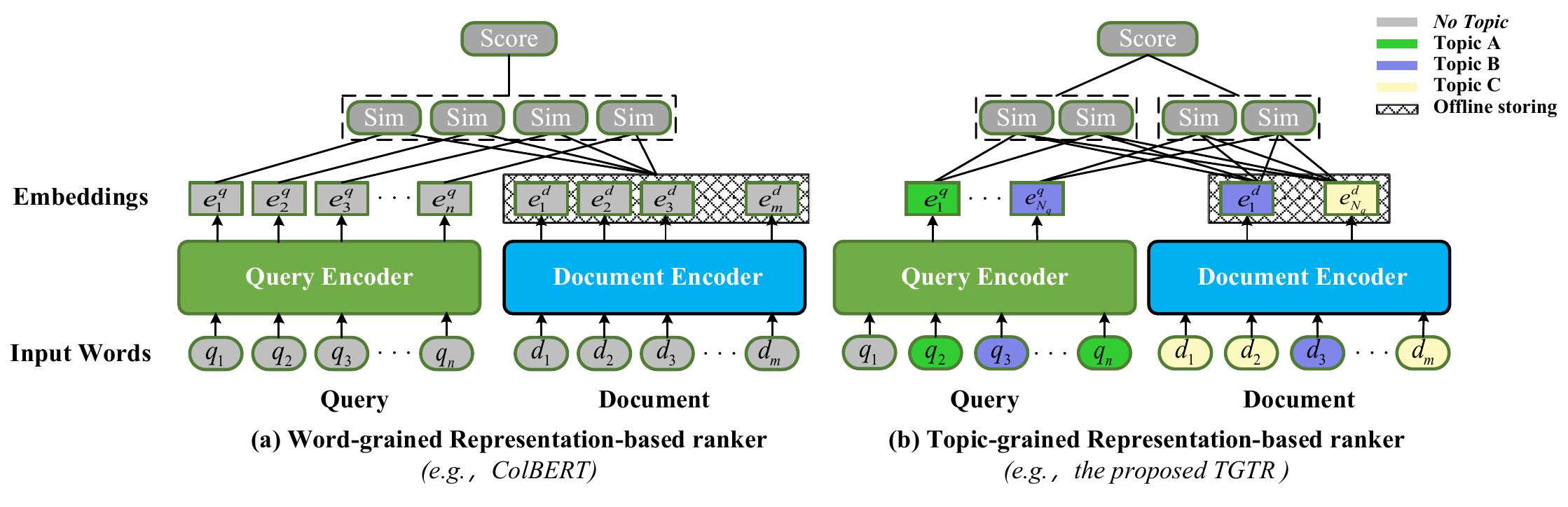}\caption{The matching paradigms of word-grained retriever (a) and the proposed method (b). Given a query and a document, (a) and (b) encode them into word-level embeddings and topic-level embeddings, respectively. (b) reduce the length of document's representation from the word count level to the topic count level,  which means it significantly compresses the space cost of offline storing documents' representations.} \label{fig1}
\end{figure}

In order to improve the retrieval speed, researchers propose a representation-based framework, where they encode query and document into word-grained representations \cite{colbert}, as shown in Fig. \ref{fig1}(a). And then they assess the matching degree of a query and a document pair by calculating the similarity of their representations. Benefit from the decoupling computation of queries and documents, the representation-based framework can pre-store document representations offline. Thus the online retrieval only needs to encode the query while it obtains the document representations from local storage directly.
However, the representation-based framework come at a steep increase in space cost to store document representations. For example, when using ColBERT \cite{colbert} to generate document representations, it requires 154 GiBs to store the TREC CAR corpus and 632 GiBs to store the MS MARCO corpus, where the document sizes of two corpora are only 2.9 GiBs and 15.6 GiBs, respectively. In this paper, we explore a novel method to compress document representations, with the goal of saving storage space and guaranteeing the retrieval accuracy as well.
\vspace{-5mm}
\vspace{-5mm}

To address the above issues, we proposes TGTR, a topic-grained text representation-based document retrieval model, as shown in Fig. 1(b). To be specific, we first model the topics distribution of documents and queries to obtain every word's latent topics, and then use Attention network to obtain topic-level embeddings by fusing words' contextual embeddings with the same topic. The motivation is drawn from the fact that, in general, users are only interested in documents consisting of information closely related to their search topic. The information in a document may cover multiple topics, and users tend to pay more attention to those parts of the document which are closely related to the query topic and less attention to the remainder. Unfortunately, previous document retrieval models \cite{DPR,COIL,colbert,PROP,RocketQA} ignore to take the topic information into account. In addition, the problem is particularly acute for long documents. 

We see the following intuitive benefits when using topic-grained representations of queries and documents to retrieval documents.
\begin{itemize}
  \item [1)] 
  Compresse the space cost of storing document representations. Compared to word-grained retrievers, the proposed topic-grained retriever can compress the size of the document representations by one order-of-magnitude, and the compression rate increases as the document length increases.
  \item [2)]
  Keep a balance of the amount of information between each embedding in query and document's  representations. We think it's the main reason why TGTR can achieve better retrieval accuray than the alternative methods described in Section 2.
  \item [3)]
  Break the existing information fusion that follows the structure of the article. We fuse the contents sharing the same topic together across the whole article. It's a process of distilling the representive information of a document. Furthermore, we find that no-topic words are frequently filler words. Which indicates our model is effective at filtering out redundant information.
\end{itemize}
In summary, this work makes the following contributions:
\begin{itemize}
  \item [1)] 
  We propose a novel document retrieval model that introduces topic-grained representation to the task for the first time;  
  \item [2)]
  Our model guarantees retrieval accuracy while significantly compressing the storage space of the document representations;
  \item [3)]
  Our model obtains competitive performance compared to all baselines on two benchmark datasets in terms of retrieval accuracy, but it requires less than 1/10 of the space cost compared to them.
\end{itemize}

\section{Related Work}\label{related}
Classical information retrieval (IR) systems rely on exact lexical match \cite{BM25}, we call them lexical retrievers. Lexical retrievers can process queries very quickly. Nowadays, they are still widely used in production systems \cite{COIL}. Recently, researchers have utilized deep learning to improve traditional lexical retrievers, including document expansions \cite{doc2query, docTTTTTquery}, query expansions \cite{query_exp} and term weight estimation \cite{deepct}.

In the past few years, information retrieval researchers have introduced a range of neural models for semantic retrieval \cite{bert-qe,COIL,colbert,condenser,DPR,DRMM,PROP,RocketQA}. Due to the specific requirements of time efficiency, researchers proposed the representation-based retrieval framework. \cite{bert,condenser,COIL,DPR,colbert,seed-encoder,RocketQA,SNRM}.

Khattab. Omar et al. \cite{colbert} first proposed a word-grained representation-based retriever, we call this type of models word-grained retrievers. Word-grained retrievers provides state-of-the-art performance at that time while resulting in significant storage overhead. COIL \cite{COIL} is another word-grained retriever which stores the token embedding in an inverted list. Representing queries and/or documents separately with a single embedding is an important method to compress document representations, which we call global-grained retrievers \cite{condenser,DPR, RocketQA}

Global-grained retrievers also generate word-level embeddings firstly, but then they fuse the sequence of embeddings into one by various means. Sentence-BERT \cite{sentence-bert} explores the effect of using 1) [CLS] embedding; 2) average pooling; 3) max pooling to fuse the BERT embedding sequence, respectively. However, This type of models can seriously impair retrieval accuracy. we attribute it to that the amount of information between query and document in the real world is often asymmetric ($|query| \ll |document|$), which leads to an imbalance of the amount of information between each embedding in query and document's representations.

In summary, current representation-based retrievers face the tradeoff of the space cost (document representations) and retrieval accuracy. TGTR effectively reduces the cost of space by constructing the topic-grained representation, without compromising retrieval accuracy.

\section{TGTR}
In this section, we present our topic-grained text representation-based model for document retrieval. Before we present the framework, some preliminary about representation-based matching paradigm are introduced. Then the TGTR framework are described in detail.

\subsection{Preliminary}
In the field of document retrieving, specially for deep models, it's very common to assess the matching degree of a query-document pair by representing the query and/or document as a sequence of vectors which we called representation-based matching paradigm. Given a query sequence $Q = [q_1, q_2, ..., q_n]$ and a document sequence $D = [d_1, d_2, ..., d_m]$, both $q_i$ and $d_j$ represent a word. Firstly, encoding a query and document into representations $E^q$ and $E^d$, then calculating the similarity between $E^q$ and $E^d$ \cite{Multi-level}. As Fig. \ref{fig1}(a) shows, traditional word-grained retrievers encode every word into a fixed-length embedding $E^q = [e^q_{1}, e^q_{2}, ..., e^q_{n}]$ and $E^d = [e^d_{1}, e^d_{2}, ..., e^d_{m}]$.

By design, the representation-based matching paradigm isolates almost all of the computations between queries and documents to enable pre-computing document representations offline \cite{colbert}. It proceeds over the documents in the collection in batches, once the documents' representations are produced, they are saved to disk using 32-bit or 16-bit values to represent each dimension. In Fig. \ref{fig1}, we use the rectangular box with decorative pattern to identify the part of offline storing. 

Generally, per embedding in representation is about hundreds of dimensions and storing a dimension needs at least 16-bit. The number of a document's embeddings stored by word-grained retrievers is approximately equal to the document length, making huge space cost. As Fig. \ref{fig1}(b) shows, we propose to encode the document with $N_d$ representative topics into topic-grained representations $E^{d} = [e^d_{_1}, e^d_{2}, ..., e^d_{N_d}]$ rather the traditional word-grained representations. The idea’s purpose is to reduce the number of the document's embeddings to compress the space cost of storing documents' representations.

\subsection{Model Architecture}
\begin{figure}[h]
\centering
\includegraphics[width=\textwidth]{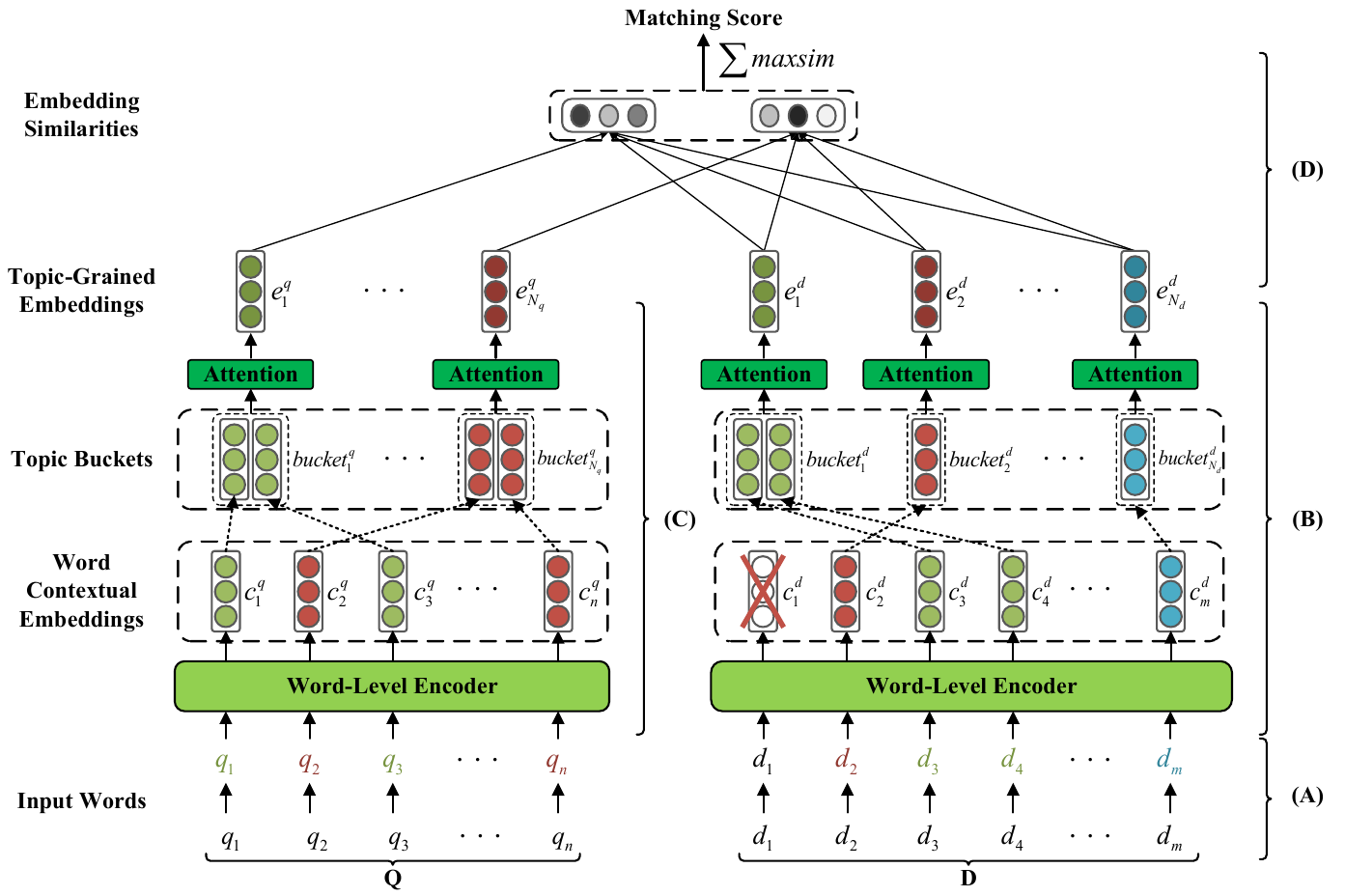}
\caption{The architecture of TGTR. (A) Topic Recognizer; (B) Query Encoder; (C) Document Encoder; and (D) Matching Assessment Mechanism. Given an input query sentence $Q$ and a document sentence $D$, (A) recognizes latent topics of every word in documents and queries, and then (B) and (C) encode the query and the document to sequences of topic-grained embeddings by three stages, separately. Finally, (D) uses maximum similarity operation to output final matching score between the query and the document.} \label{fig2}
\end{figure}

Fig. \ref{fig2} depicts the architecture of TGTR, which comprises four components. We will cover these components in detail in this section.
\vspace{-5mm}
\subsubsection{Topic Recognizer}\label{Topic}
The topic recognizer uses traditional topic generation model to recognize latent topics of every word in documents and queries. As Fig. \ref{fig2}(A) shows, given a query sequence $Q = [q_1, q_2, ..., q_n]$ and a document sequence $D = [d_1, d_2, ..., d_m]$, topic recognizer gives their words different topic colors by analyzing their topics distributions.\\
\indent To be specific, we use Latent Dirichlet Allocation (LDA) \cite{LDA} to model documents and queries' latent topics. Algorithm \ref{alg:topic} depicts the process of obtaining words' topics. Firstly, we obtain the text-topic distribution Array as well as the topic-word distribution Array (line 1-2). We then set threshold $\theta_{t}$ to extract the representative topics for each text (line 3-8). In the same manner, we set fixed threshold $\theta_{wf}$ and ratio threshold $\theta_{wr}$ to extract the representative words in a text under its representative topics (line 9-11). Finally, a two-dimensional table $M$ is generated for each text (line 12). The dim of row in $M$ represents a word, while the dim of column represents a potential topic. Note a word may have more than one or zero latent topics. The words without any latent topics are considered meaningless and discarded after helping other words to generate contextual embeddings (see the red cross symbol in Fig. \ref{fig2}).
\vspace{-5mm}
\begin{algorithm}[htb] 
\caption{Topic Recognizer of TGTR} \label{alg:topic}
\hspace*{0.02in} {\bf Input:}\\
\hspace*{0.15in} The text-words array, ${\rm text2words}$;\\
\hspace*{0.15in} The number of texts, $m$;\\
\hspace*{0.15in} The number of topics, $k$;\\
\hspace*{0.15in} The trained topic model, ${\rm LDA}$;\\
\hspace*{0.15in} The word frequency builder, ${\rm Vectorizer}$;\\
\hspace*{0.15in} The threshold for extracting representative topics , $\theta_{t}$;\\
\hspace*{0.15in} The fixed and ratio threshold for extracting representative words, $\theta_{wf}$, $\theta_{wr}$;\\
\hspace*{0.02in} {\bf Output:}\\ 
\hspace*{0.15in} The list of potential topics of words in all texts, $\rm text2word2topics$;
\begin{algorithmic}[1]
\State ${\rm textVectorizer} \gets {\rm Vectorizer.transform}({\rm text2words})$
\State ${\rm text2topics, topic2words} \gets {\rm LDA.transform}(\rm{textVectorizer})$
\For{$d \in [1,2,...,m]$}
    \State $M \gets [\;]$, ${\rm text2word2topics} \gets [\;]$
    \For{$t \in [1,2,..., k]$}
        \If{${\rm text2topics}{[d, t]} \ge \theta_{t}$}
            \State Get the distribution of words in the $d$-th text \\
               \qquad \qquad \qquad \qquad under the $t$-th topic ${\rm dis}_{dt}$ $\gets$ topic2words
           
            \State $l \gets |{\rm dis}_{dt}|$
            \For{$w \in [1,2,...,l]$}
                \If{
            ${\rm dis}_{dt}[w] \ge \theta_{wf} \; \textbf{or} \;{{\rm dis}_{dt}}.{\rm getorder}(w)/l \le \theta_{wr}$}
                    \State $M[w].{\rm append}(t)$
                \EndIf
            \EndFor
        \EndIf
    \EndFor
    \State ${\rm text2word2topics}.{\rm append}(M)$
\EndFor
\State \Return $\rm text2word2topics$
\end{algorithmic}
\end{algorithm}
\vspace{-10mm}
\subsubsection{Document Encoder}
We then encode the document to a sequence of fixed-length embeddings. This part comprises three stages.
\paragraph{Stage 1: Encode word-level contextual embeddings} Given a document $D$, TGTR first maps each word $d_i$ into its contextual embedding $c_{i}^{d}$ by using Word-Level Encoder ({WLE}). Though we can complete this part of work by using methods such as in \cite{wordBert, bert}, We focus on BERT \cite{bert} to keep consistent with the major baseline. Note BERT uses WordPiece embeddings with a 30,000 token vocabulary, thus a word can be tokenized to several tokens. Strictly speaking, the $i$-th word's contextual embedding $c_{i}^{d}$ may comprises more than one token embedding, which we hope readers will notice. The process of this stage is summarized as Equation \ref{con:stage1}.
\begin{equation}
    [{c_{1}^d}, {c_{2}^d}, ..., {c_{m}^d}] := {\rm WLE}(d_1, d_2, ..., d_m) \label{con:stage1}
\end{equation}
\paragraph{Stage 2: Word-topic mapping} As mentioned above, we obtain a two-dimensional word-topics table $M$ for every document by modeling latent topics. The dim of row in $M$ represents a word, while the dim of column represents a potential topic. Every word's contextual embedding obtained in Stage 1 is mapped to the buckets corresponding to topics they have. A word may be mapped into multiple buckets or filtered out (regarded as meaningless). The bucket corresponding to the $i$-th topic is marked as $bucket_{i}^{d}$. The process of this stage is summarized as Equation \ref{con:stage2}, where $N_d$ is the number of representative topics the document $d$ has.
\begin{equation}
    [bucket_{1}^{d},bucket_{2}^{d},...,bucket_{N_d}^d] := {\rm Mapping}({c}_{1}^d, {c}_{2}^d, ..., {c}_{m}^d) \label{con:stage2}
\end{equation}
\paragraph{Stage 3: Generate topic-grained representation} The model TGTR uses \textbf{Attention} network to obtain topic-level embeddings, which we call topic-grained representation. Considering different words with the same topic have different amount of information, we assign different weights to different words. For the bucket corresponding to the $t$-th topic $bucket_{t}^d : [u_1, u_2, ...,u_{B_t}]$ outputed by stage 2, denote the attention weight of $u_i$ as $\alpha_i$:
\begin{equation}
\alpha_{i} = {q_t}^{T} \tanh(W \times {u}_{i} + {b})\label{con:attention_1}
\end{equation}
\begin{equation}
\alpha_{i} = \frac{\exp{(\alpha_{i})}}{\sum_{j=1}^{B_t}\exp{(\alpha_{j})}}\label{con:attention_2}
\end{equation}
where $W$ and $b$ are parameters, ${q}_{t}$ is the attention query vector, $tanh$ is the activation function and ${B}_t$ is the size of the $t$-th topic bucket. The final embedding of $t$-th topic $e_t^d$ is the summation of the word-level embeddings in $bucket_{t}^d$ weighted by their attentions.
\begin{equation}
{e}_{t}^{d} = \sum_{i=1}^{B_{t}}\alpha_{i}{u_i}
\label{con:u}
\end{equation}
\subsubsection{Query Encoder} Our query encoder has a very similar architecture with document encoder, they share model parameters but have a few difference in input processing. We prepend BERT’s start token [CLS] followed by a special token [D] when input a document sequence. In the same manner, we prepend BERT’s start token [CLS] followed by a special token [Q] when input a query sequence. 
\subsubsection{Matching Assessment Mechanism} Finally, we use a maximum similarity (MaxSim) operation to output our final matching score. Given the query's topic-grained representation $E^{q} : [e_{1}^q, e_{2}^q, ..., e_{N_q}^q]$ and the document's topic-grained representation $E^{d} : [e_{1}^d, e_{2}^d, ..., e_{N_d}^d]$, the matching score of query $q$ and document $d$ is assessed by MaxSim operation between $E^q$ and $E^d$. To be specific, we applies MaxSim between one of the query embeddings and all of the document’s embeddings, then we sum all items up as final score $S(Q, D)$. The process of this part is summarized as Equation \ref{con:fscore}.
\begin{equation}
    S(Q, D)=\sum_{i \in [|E^q|]}{\rm max}_{j \in [|E^d|]}e_{i}^{q}(e_{j}^{d})^T\label{con:fscore}
\end{equation}
 Notice our Matching Assessment Mechanism has no trainable parameters. 
 \subsubsection{Training}
The training objective is to learn representations of queries and documents so that query-positive document pairs have higher matching score than the query-negative documents pairs in training data. Given a query $Q$ together with its positive documents ${D}^+$ and $m$ negative documents. $\{D_i^{-}\}_{i=1}^m$, we minimize the loss function:\\
 \begin{equation}
    L(Q, D^{+}, \{D_i^{-}\}_{i=1}^m) = -{\rm log}\frac{\exp({S(Q, D^{+})})}{\exp({S(Q, D^{+})}) + \sum_{i=1}^{m}\exp({S(Q, D_i^{-})})}\label{con:loss}
\end{equation}

\section{Experiment Methodology}
\subsection{Datasets}
Following previous work \cite{colbert}, our experiments use two datasets, which differ in data size, to evaluate our model in document retrieving tasks.\\
\textbf{TREC CAR}. TREC CAR is introduced by Dietz et al. \cite{TREC}  in 2017, is a composite data set based on Wikipedia containing approximately 29 million articles. Our assessment was performed on the test set used in TREC 2017
CAR, which contained 2,254 queries.\\
\textbf{MS MARCO.} MS MARCO \cite{msmarco} is a dataset introduced by Microso
in 2016 for reading comprehension and adapted in 2018 for retrieval. It is
a collection of 8.8M passages from Web pages, which were gathered from
Bing’s results to 1M real-world queries.

\subsection{Baseline Methods}
We adopt three types of baselines for comparison.\\
\textbf{Lexical Retriever}. Lexical Retriever retrieve document based on lexical matching rather than semantic matching. In this type, we choose three traditional methods \cite{QL,BM25,TextRank} and three network methods \cite{doc2query,docTTTTTquery,deepct} as our baselines.\\
\textbf{Global-grained Retriever}. Global-grained retriever retrieve document with global-grained representations of queries and documents. In this type, we choose BERT \cite{bert} and DPR \cite{DPR} as our baselines.\\
\textbf{Word-grained Retriever}. Global-grained retriever retrieve document with word-grained representations of queries and documents. In this type, we choose ColBERT \cite{colbert} and COIL \cite{COIL} as our baselines.

\section{Experiment Details}
\subsection{Implementation Details}
The complete training details are given below:
\begin{itemize}
    \item 
    We fit LDA model by using Scikit-learn machine learning library \cite{sklearn}. We apply variational inference with expectation-maximization to learn model's parameters and get the distributions described in Section 3. The number of latent topics $K$ is a hyper-parameters here, and we set other two hyper-parameters $\alpha$ and $\eta$ to $1/K$ by default.
    \item 
    We choose the max query length as 32 and the max doc length as 180 at dataset MS MARCO. Since TREC is much larger than MS MARCO, we set max query length 48 and max doc length 250 in TREC.
   \item 
   We use BERT as pre-trained word-level embedding encoder to embed the query and document sentences with the embedding dimension of 768 and the vocab size of 30522.
    \item 
    We then apply attention operation for every topic buckets by different query vectors. The parameters $W$ and $b$ are shared by all buckets.
    \item  
    The dimension of final topic-level embedding $dim$ is 768. We passes the embeddings through a linear layer with no activations to control their dimensions. As we discuss later in more detail, we typically fix $dim$ range as (64, 128, 256, 512, 768). We set $dim$=$256$ by default.
\end{itemize}

\subsection{Experiment Results}
Table \ref{tab:main} shows the retrieving performance of TGTR and our baselines over two datasets.\\
\indent\textbf{Compared to word-grained retrievers} The results show that TGTR performs almost 10 times better than the word-grained baselines in terms of space cost with no loss in terms of retrieval accuracy on MS MARCO. On TREC CAR, TGTR performs almost 12 times better than the word-grained baselines in terms of space cost with 2.3\% and 2.0\% loss in terms of MRR@10 and MAP on MS MARCO.\\
\indent\textbf{Compared to global-grained retrievers} The results show that TGTR overwhelmingly outperforms global-grained retrievers in terms of space cost and retrieval accuracy over both datasets. Note our model outperforms global-grained retrievers in terms of space cost because we reduce the embedding dimension by passing the original embeddings through a linear layer.\\
\indent\textbf{Compared to lexical retrievers} The results show that TGTR overwhelmingly outperforms lexical retrievers in terms of retrieval accuracy over both datasets. Note lexical retrievers don't need store documents' representations, so the compare between our model and them in terms of space cost is not available.\\
\indent\textbf{Summary.} Compared to word-grained baselines, TGTR is consistently outperforming them on MS MARCO and be competitive with them on TREC CAR in terms of retrieval accuracy, but it performs almost 10 times better than them in terms of space cost. Moreover, TGTR overwhelmingly outperforms lexical and global-grained baselines in terms of retrieval accuracy.
\vspace{-4mm}
\begin{table}[H]
\caption{Retrieving performances of TGTR and baseline models. We report the performances of our model with the embedding dimension $dim$=$256$. Improvement, degradation or equivalent with respect to TGTR in terms of MRR@10, Recall@1K and MAP is indicated (+/$-$/-). The unit of `Space' is (GiBs). Results not applicable are denoted `n.a.’.}\label{tab:main}
\centering
\subtable{(a) Performance Comparisons on MS MARCO.}{
\resizebox{!}{!}{
\begin{tabular}{l|cc|cc|cc}
\hline
{\bfseries{Method}}&\multicolumn{2}{c}{\bfseries{Space(GiBs)}}\vline&\multicolumn{2}{c}{\bfseries{MRR@10}}\vline&\multicolumn{2}{c}{\bfseries{Recall@1K}}\\
\cline{1-7}
{Lexical Retriever}&&&&&&\\
BM25&		 n.a.& n.a.& 0.187&$-$48.2\%&0.857&$-$11.5\%\\
Doc2query&	 n.a. &n.a. &0.215 &$-$40.4\% &0.891 &$-$8.0\%\\
DeepCT&    n.a.&n.a. &0.243 &$-$32.7\% &0.910 &$-$6.0\%\\
DocTTTTTquery& n.a. &n.a. &0.277 &$-$23.3\% &0.947 &$-$2.2\%\\
\hline
{Global-grained Retriever}&&&&&\\
BERT&  25.3 &×1.6 &0.310 &$-$14.1\% &0.929 &$-$4.0\%\\
DPR & n.a. &n.a. &0.311 &$-$13.9\% &0.952 &$-$1.7\%\\
\hline
{Word-grained Retriever}&&&&&\\
COIL&  n.a. &n.a. &0.355 &$-$1.7\% &0.963 &$-$0.5\%\\
ColBERT& 154.0 &×9.9 &0.360 &-0.3\% &\bfseries{0.968} &\bfseries{-}\\
\hline
{Topic-grained Retriever}&&&&&\\
TGTR&	\bfseries{15.6} &\bfseries{×1} &\bfseries{0.361} &\bfseries{-} &\bfseries{0.968} &\bfseries{-}\\
\hline
\end{tabular}
}
\label{firsttable}
}
\\
\vspace{3mm}
\subtable{(b) Performance Comparisons on TREC CAR.}{
\resizebox{!}{!}{
\begin{tabular}{l|cc|cc|cc}
\hline
{\bfseries{Method}}&\multicolumn{2}{c}{\bfseries{Space(GiBs)}}\vline&\multicolumn{2}{c}{\bfseries{MRR@10}}\vline&\multicolumn{2}{c}{\bfseries{MAP}}\\
\cline{1-7}
{Lexical Retriever}&&&&&&\\
BM25&		 n.a.& n.a.&  n.a.& n.a.&0.153&$-$50.2\%\\
TextRank&	 n.a.& n.a.& 0.160&$-$63.0\%&0.120&$-$60.9\%\\
Doc2query&	 n.a. &n.a. & n.a. & n.a. &0.181 &$-$41.0\%\\
DeepCT&    n.a.&n.a. &0.332 &$-$23.3\% &0.246 &$-$19.9\%\\
\hline
{Global-grained Retriever}&&&&&\\
BERT&  83.2 &×1.6 &0.376 &$-$13.2\% &0.273 &$-$11.1\%\\
\hline
{Word-grained Retriever}&&&&&\\
ColBERT& 632.1 &×12.3 &\bfseries{0.443} &+\bfseries{2.3}\% &\bfseries{0.313} &+\bfseries{2.0\%}\\
\hline
{Topic-grained Retriever}&&&&&\\
TGTR&	\bfseries{51.4} &\bfseries{×1.0} &0.433 &- &0.307 &-\\
\hline
\end{tabular}
}
\label{secondtable}
}
\end{table}
\vspace{-10mm}

\section{Analysis}
\subsection{A Comparison of Trade-off Quality}
In this section, we assess the trade-off quality between space efficiency and retrieval accuracy of three types of representation-based retrievers. We use the quotient of MRR and Space as the trade-off score. We use BERT and ColBERT to represent the global-grained retriever and word-grained retriever, separately. Topic-grained Retriever is our model. 

Table \ref{tab:tradeoff} shows the results. It seems that our model significantly outperforms other two types of retrievers in term of trade-off quality and the word-grained retriever performs the worst.
\vspace{-5mm}
\begin{table}[H]
\caption{Comparisons of Trade-off Quality among Three Types of Retrievers.}\label{tab:tradeoff} 
\centering
\resizebox{!}{!}{
\begin{tabular}{l|cc|cc}
\hline
\multirow{2}*{\bfseries{Method}}&\multicolumn{4}{c}{\bfseries{MRR/Space} (1e-3)}\\
\cline{2-5}
~&\multicolumn{2}{c}{\bfseries{\;MS MARCO\;}}\vline&\multicolumn{2}{c}{\bfseries{\;TREC CAR\;}}\\
\hline
Global-grained Retriever & 12.3 &$-$46.8\% & 4.5 &$-$46.4\%\\
Word-grained Retriever &2.3 &$-$90.0\% & 0.7 &$-$91.7\%\\
\hline
Topic-grained Retriever & \textbf{23.1} &\textbf{-} & \textbf{8.4} &\textbf{-}\\
\hline
\end{tabular}
}
\end{table}
\vspace{-10mm}

\subsection{Embeddings Dimension and Bytes per Dimension}
\begin{figure}[h]
\centering
\includegraphics[width=\textwidth]{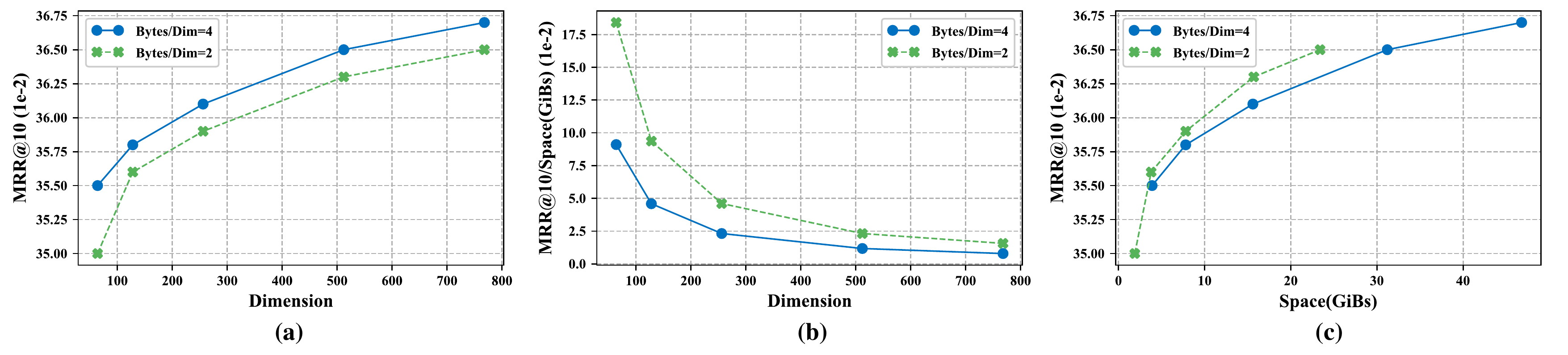}
\caption{(a) and (b) show the impact of embeddings dimension and the bytes per dimension on MRR@10 and trade-off quality (MRR/Space), separately. (c) shows MRR@10 vs Space(GiBs) as functions of the embeddings dimension and the bytes per dimension.} \label{fig5}
\end{figure}

Two of the most attractive features in our model is the embeddings dimension and the bytes per dimension. Fig. \ref{fig5} shows the impact of above two features on the model performance.
\vspace{-5mm}
\vspace{-5mm}
As Fig. \ref{fig5}(a) shows, retrieval accuracy increases sublinearly with the increase of above two features in our model. Fig. \ref{fig5}(b) clearly shows that it might contribute to higher trade-off quality by reducing the above two features. As Fig. \ref{fig5}(c) shows, retrieval accuracy increases sublinearly with the increase of space cost in our model. It seems that when the embedding dimension is small enough, further compression can cause great accuracy damage.

\section{Conclusions}
This paper presents TGTR, a novel retrieval model that employs topic-grained text representation for document retrieval. The key of our model is modeling the topics distribution of documents and queries to obtain every word’s latent topics, and then using Attention network to obtain topic-level embeddings by fusing words’ contextual embeddings with the same topic. Our experiments on MS MARCO and TREC benchmark datasets demonstrates the advantage of representing texts as topic-grained embeddings for document retrieval task. These results suggest that our model guarantees retrieval accuracy while significantly compressing the storage space of the document representations.
\bibliographystyle{splncs04}
\bibliography{ref}
\end{document}